\documentclass[pra,superscriptaddress,aps,nofootinbib]{revtex4}

\usepackage{amsmath,epsfig,amsfonts,graphicx,hyperref}
\usepackage{subfigure}
\usepackage{graphicx}
\usepackage[american]{babel}
\usepackage{amssymb}
\usepackage{amsfonts}
\usepackage{color}
\usepackage{comment}
\usepackage{marvosym}
\usepackage{ifsym}
\usepackage[normalem]{ulem}

\begin{document}
	\newcommand{\be}{\begin{equation}}
	\newcommand{\ee}{\end{equation}}
	\newtheorem{corollary}{Corollary}[section]
	\newtheorem{remark}{Remark}[section]
	\newtheorem{definition}{Definition}[section]
	\newtheorem{theorem}{Theorem}[section]
	\newtheorem{proposition}{Proposition}[section]
	\newtheorem{lemma}{Lemma}[section]
	\newtheorem{help1}{Example}[section]

\def\re{\text{Re}}
\def\im{\text{Im}}
\def\labelitemi{$\blacktriangleright$}

\title{Excitation of Peregrine-type waveforms from vanishing initial conditions in the presence of  periodic forcing}

\author{Nikos I. Karachalios\footnote{\Letter\;\; Corresponding Author.
		 Email: karan@aegean.gr. ORCID ID: \url{https://orcid.org/0000-0002-5580-3957}}}, 
\affiliation{Department of Mathematics, University of the Aegean, Karlovassi, GR 83200
	Samos, Greece}
\author{Paris Kyriazopoulos}
\affiliation{Department of Mathematics, University of the Aegean, Karlovassi, GR 83200
	Samos, Greece}
\author{Konstantinos Vetas}
\affiliation{Department of Mathematics, University of the Aegean, Karlovassi,  GR 83200
	Samos, Greece}

\begin{abstract}
We show by direct numerical simulations that spatiotemporally localized wave forms, strongly reminiscent of the Peregrine rogue wave, can be excited by vanishing initial conditions for the periodically driven nonlinear Schr\"odinger equation. The emergence of the Peregrine-type waveforms can be potentially justified, in terms of the existence and modulational instability of spatially homogeneous solutions of the model, and the continuous dependence of the localized initial data for small time intervals. We also comment on the persistence of the above dynamics, under the presence of small damping effects, and justify, that this behavior should be considered as far from approximations of the corresponding integrable limit. 
\end{abstract}

\maketitle

\section{Introduction  }
A crucial and intriguing question in nonlinear dynamics, concerns the persistence of dynamical features of integrable systems in the presence of perturbations. In the context of the nonlinear Schr\"odinger (NLS)  partial differential equations, and particularly for the integrable NLS (with a focusing, cubic nonlinearity), one of these features which is receiving tremendous interest, concerns the emergence of rogue waves: extreme wave events possessing spatiotemporal localization and large amplitude, which are mathematically described by its class of rational solutions.  For the fundamental representatives of this class, namely the Peregrine rogue wave (PRW), and the
space- or time-periodic solutions as the Akhmediev and Kuznetsov-Ma
(KMb) breathers, respectively, \cite{H_Peregrine},\cite{kuz},\cite{ma},\cite{akh}, their physical relevance is justified by numerous episodes in the ocean \cite{k2a},\cite{k2b},\cite{k2c}, and experimental observations in hydrodynamics \cite{k2d},\cite{hydro},\cite{hydro2}, nonlinear optics and lasers \cite{opt1},\cite{opt2},\cite{laser}, superfluidity \cite{He}, and  plasma physics \cite{plasma}. Such natural and experimental evidences, motivated recent advances on the predictability of extreme wave events, based on studies analyzing the interactions of energy localization and strong local nonlinearity \cite{Sapsis1},\cite{Sapsis2}.

In the context of the aforementioned  persistence question, the robustness of rational solutions and rogue wave dynamics in the presence of perturbations, has been identified for various special cases of extended NLS equations. Representative key works, refer to third order (including modified Hirota \cite{Hirotar}, and Dysthe \cite{dt}, equations) \cite{devine}, \cite{calinibook}, \cite{NRbor5a}, as well as, fourth \cite{ NR5Wang},\cite{NR4Anki}, and fifth order \cite{NRbor6}, models.  Important extensions to coupled equations and systems include \cite{BorPT} for  parity-time symmetric systems, \cite{BorCD} for NLS systems with derivative nonlinearities, and \cite{BorMAN} for Manakov systems (physically relevant in the context of Bose-Einstein condensates).  In a different perspective, results on the spectral analysis of the PRW as a limiting case of the KMb, are presented in \cite{All1}. Furthermore,  the findings of \cite{BorDPJ} suggest that a suitably defined dispersion or nonlinearity management, if applied to a continuous wave (cw) background, may effectively stabilize the supported PRW and KMb waveforms. 
%

In this spirit, and motivated by key works on the linearly forced/driven NLS equations in the context of rogue waves and the robustness of localized waveforms in nearly integrable systems \cite{EPeli},\cite{Kharif1},\cite{Kharif2},\cite{BoYu}, we consider instead of the linear forcing, the action of an external, time-periodic driver. The relevance of such a driving term, arising in a number of physical systems (as in the theory of charge-density waves and plasma physics) has been extensively analyzed in \cite{BoYu}. Then, continuing along the lines of our recent work \cite{All2}, we examine,  starting herein numerically, the potential {\it emergence of spatiotemporal algebraically decaying waveforms in the dynamics of the associated, periodically driven NLS model}. 
	
However, the approach we will follow regarding the initial conditions, differs drastically from investigations exciting PRWs, from initial data defined as interactions between a pulse of small amplitude and a cw, \cite{All2},\cite{BM},\cite{Yang1},\cite{Yang2}. Instead of such initial conditions, we ask for the possibility to excite PRWs {\em from vanishing initial conditions}, decaying either at an algebraic, or exponential rate. 
 
The answer is positive: the main finding is that extreme wave forms, strongly reminiscent to the PRW, can be excited from the decaying initial data.  Their profile and statiotemporal decay rates are close to that of the analytical PRW solution. An important feature of this finding is that these  PRW-solitonic structures emerge on the top of a finite background, which is formed at the early stages of the evolution, although the initial condition decays to zero.  The birth of the PRW-type waveforms can be potentially understood, by a synergy of the modulational instability of the cw-solutions of the model, and the persistence of the localization of the initial condition on the top of the emerged unstable background. In other words, for vanishing initial data, in the presence of the periodic forcing, the system self-induces the effects of the modulational instability mechanism analyzed in \cite{All2},\cite{BM},\cite{Yang1},\cite{Yang2}, for the excitation of PRWs. 

The paper is structured as follows:
In Section~\ref{numerical}, we report the results of the numerical simulations and analytical considerations on the stability of spatial homogeneous solutions. We also briefly comment on the dynamics of the linearly damped and forced model, and of the integrable  NLS, initiated from the same, decaying initial conditions. In Section~\ref{conclusions}, we summarize our results with an eye towards future work.
\section{Numerical Results}
\label{numerical}
\paragraph{ Brief description of the model.}
In this section, we report the results of direct numerical simulations, on the dynamics of the periodically driven, nonlinear Schr\"odinger (NLS) equation  
\begin{eqnarray}
\label{eq1}
\mathrm{i}{{u}_{t}}+\frac{\nu}{2}{{u}_{xx}}+\sigma|u|^2u =\Gamma\exp(\mathrm{i}\Omega t),\;\;\nu>0,\;\sigma>0,\;\Gamma\in\mathbb{C},\;\;\Omega\in\mathbb{R}.
\end{eqnarray}
%
In  Eq.~ (\ref{eq1}), the parameter $\nu$ is the second order (group velocity) dispersion, and  $\sigma$ is the strength of the nonlinearity. The parameters $\Gamma$ and $\Omega$, correspond to the amplitude and frequency of the driver, respectively. Let us recall, that under the change of variable $u\rightarrow u \exp(\mathrm{i}\Omega t)$, Eq.~\eqref{eq1} is transformed to the autonomous equation $\mathrm{i}{{u}_{t}}+\frac{\nu}{2}{{u}_{xx}}+(\sigma|u|^2-\Omega)u =\Gamma$. In what follows, we shall restrict to the case $\Gamma\geq 0$.\footnote{For the generic case of $\Gamma\in\mathbb{C}$, we remark that amplitude and phase may seriously affect the dynamics that  will be discussed herein. Relevant investigations are in progress, and will be considered elsewhere.}

Eq.~\eqref{eq1} defines a non-integrable  perturbation of the  focusing integrable NLS, which corresponds to the case  $\Gamma=0$.  It is one of the fundamental partial differential equations exhibiting complex\cite{Rev1}, even spatiotemporal chaotic behavior \cite{RK1},\cite{RK2}, together with its dissipative counterparts \cite{nobe1},\cite{nobe2},\cite{NB86}. The impact of the breaking of the hyperbolic structure of the integrable NLS, in the emergence of complex dynamics for the damped and forced models, has been rigorously analyzed in  \cite{Li},\cite{Wig1},\cite{kai},\cite{CLM}. For equations incorporating higher order dissipation, we refer to \cite{BoYu},\cite{RK3} and references therein.

The numerical experiments will simulate the dynamics of Eq.~\eqref{eq1}, excited by either, the quadratically decaying  initial condition
\begin{eqnarray}
\label{eq4}
u_0(x)=\frac{1}{1+x^2},
\end{eqnarray}
or the exponentially decaying
\begin{eqnarray}
\label{eq5}
u_0(x)=\alpha\mathrm{sech}(\beta x),\;\;\mbox{for some}\;\;\alpha,\;\beta>0,
\end{eqnarray}
which resembles the profile of a bright soliton solution of the integrable NLS. The model will be supplemented with periodic boundary conditions 
\begin{eqnarray}
\label{bc}
u(x+2L,t)=u(x,t),\;\;\mbox{for all}\;\;t\geq 0.
\end{eqnarray}  
\paragraph{Background.}
 The system was integrated by using a pseudo-spectral method for the spatial discretization and an adaptive step Runge-Kutta method for the time-stepping.  Concerning the implementation of the periodic boundary conditions, for the cases \eqref{eq4} and \eqref{eq5} of the initial data, let us recall the following:  These conditions are strictly satisfied only asymptotically, as $L\rightarrow\infty$. 
For a finite length $L$, the initial profiles, as well as, their spatial derivatives, 
have jumps across the end points of the interval $[-L, L]$. However, these jumps have 
negligible effects in the observed dynamics, as they are either of order $1/(1+L^2)$ or $\exp(-L)$. More precisely, 
since the smallest value for $L$ used herein, is $L=100$, the effects are of order 
$\mathcal{O}(10^{-4})$, 
or less. The characterization of the numerically observed spatiotemporal localized waveforms as extreme, is made by a comparison \cite{BS}, against time-translations of the analytical PRW solution  of the integrable NLS:
\begin{eqnarray}
u_{\mbox{\tiny PS}}(x,t;P_0)=\sqrt{P_0}\left[1-\frac{4\left(1+\frac{2\mathrm{i}t}{\Lambda}\right)}{1+\frac{4x^2}{K_0^2}+\frac{4t^2}{\Lambda^2}}\right]\exp\left(\frac{\mathrm{it}}{\Lambda}\right),
\label{sprw}
\end{eqnarray}
where the parameters $\Lambda=\frac{1}{\sigma\,P_0}$ and $K_0=\sqrt{\nu\,\Lambda}$. 
Note that the PRW solution decays algebraically both in time and space, on the top of the continuous background of power $P_0$.  We denote these translations as $u_{\mbox{\tiny PS}}(x,t-t^*;P_0)$. The time $t^*$ and power $P_0$, are numerically detected, as described below. 
\begin{figure}[tbh!]
	\hspace{-0.5cm}\includegraphics[scale=0.19]{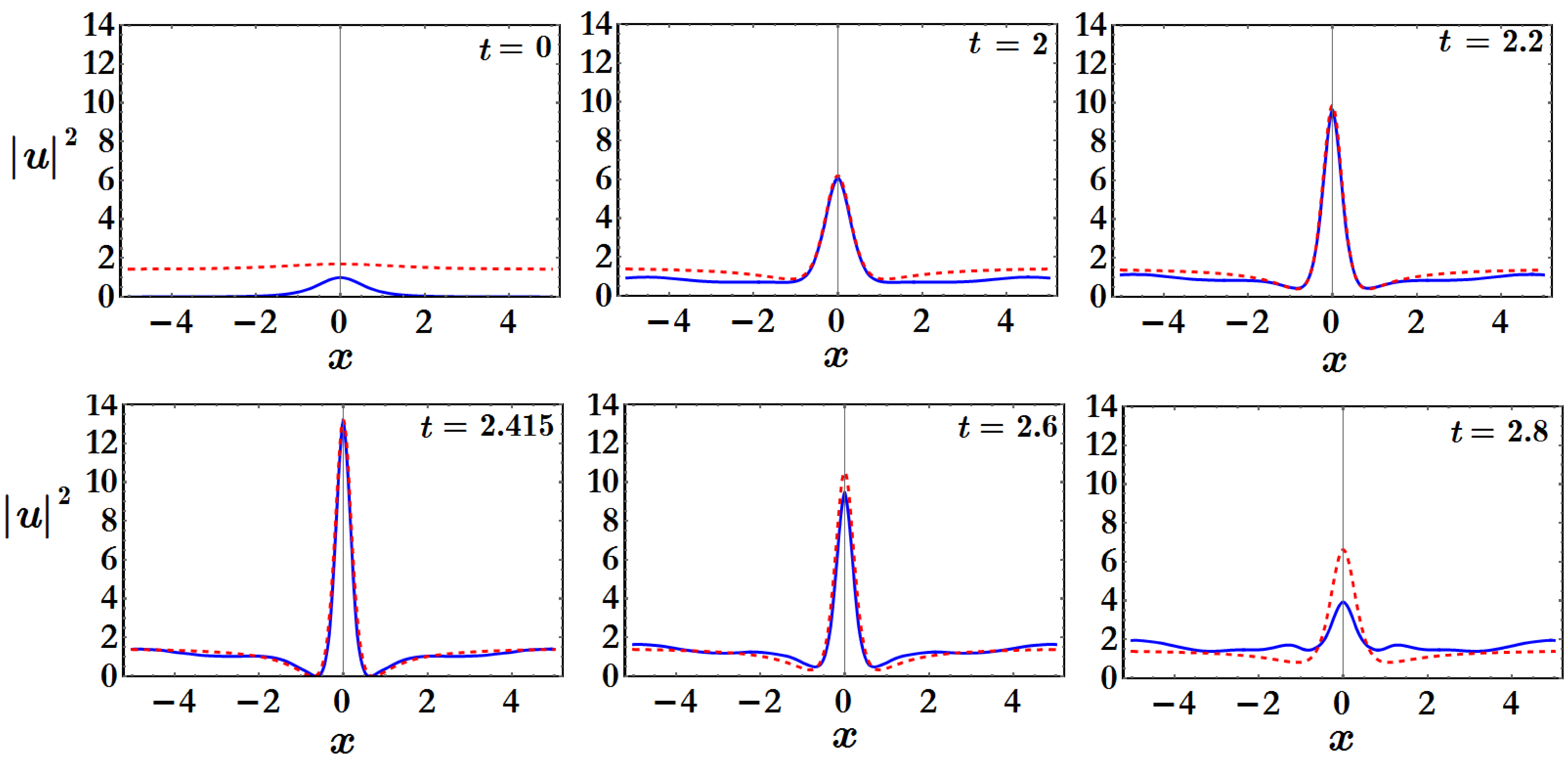}
	\caption{(Color Online) Snapshots of the evolution of the density $|u(x,t)|^2$ [solid (blue) curves], for the initial condition (\ref{eq4}). Parameters: $\nu=1$, $\sigma=1$, $\Gamma=0.5$, $\Omega=1$, $L=100$. The density of the numerical solution of Eq.~\eqref{eq1} is compared against the density of 
		the PRW of the integrable NLS (\ref{sprw}), $u_{\mbox{\tiny PS}}(x,t-2.415;0.84)$ [dashed (red) curves], with $K_0=1.54$ and $\Lambda=1.19$. 
	}
	\label{figure1}
\end{figure}
\paragraph{Results of the direct numerical simulations.}
Fig.~\ref{figure1}, shows snapshots of the evolution of the density $|u(x,t)|^2$ of the numerical solution of the problem~\eqref{eq1}-\eqref{eq4}-\eqref{bc},  for $\nu=\sigma=1$, $\Gamma=0.5$ and $\Omega=1$, for the algebraically decaying initial condition \eqref{eq4}.  We observe that the initial datum evolves towards a localized waveform,  which is strongly reminiscent of a PRW.  The numerical solution is plotted by the continuous (blue) curve, against the dashed (red) curve depicting the evolution of the PRW-profile (\ref{sprw}), $u_{\mbox{\tiny PS}}(x,t-2.415;0.84)$.  The maximum amplitude of the event is attained at $t^*=2.415$. The time $t^*$ is used to define the  time-translation of the analytical PRW-solution of the integrable NLS. The power of its background $P_0=0.84$ is numerically detected, so that the amplitude of the analytical PRW coincides with the maximum amplitude of the numerical event. Note that for the above set of parameters, we found that $K_0=1.54$ and $\Lambda=1.19$.  

\begin{figure}[tbh!]
	\hspace{-0.5cm}\includegraphics[scale=0.19]{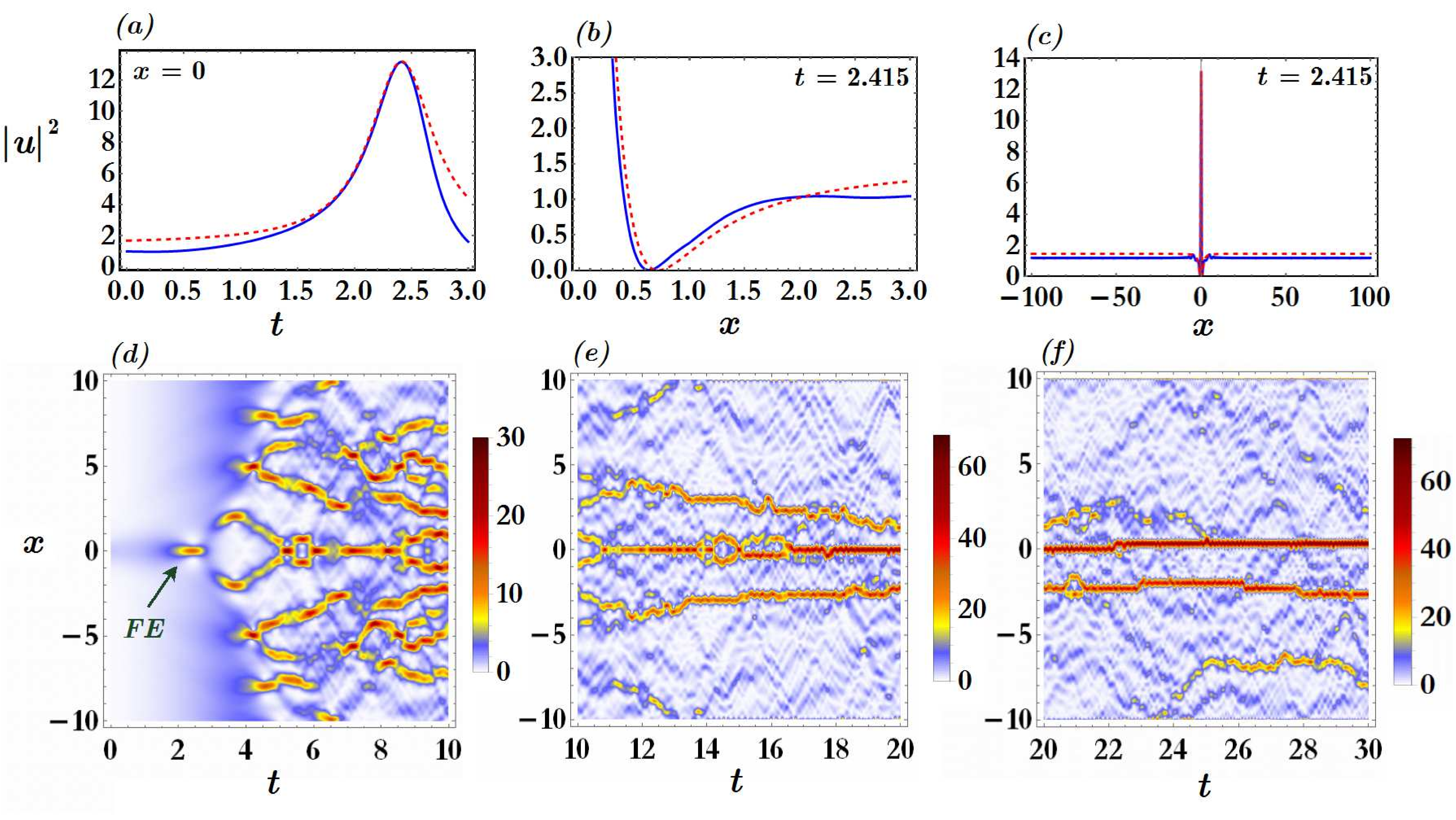}
	\caption{(Color Online) Parameters: $\nu=1$, $\sigma=1$, $\Gamma=0.5$, $\Omega=1$, $L=100$. Top left panel (a): evolution of the density of the center, $|u(0,t)|^2$, for the initial condition (\ref{eq4}), against the evolution of the density of the center of the PRW (\ref{sprw}),  $u_{\mbox{\tiny PS}}(x,t-2.415;0.84)$, with $K_0=1.54$ and $\Lambda=1.19$. Top middle panel (b): a detail of the spatial profile of the maximum event at $t^*=2.415$, close to the right of the two symmetric minima of the exact PRW  $u_{\mbox{\tiny PS}}(x,t-2.415;0.84)$. Top right panel (c): Another view of the numerical density, at time $t^*=2.415$, where the extreme event attains its maximum amplitude, for $x\in[-100,100]$. The PRW-type structure is formed on a finite background. Bottom panels: Contour plots of the spatiotemporal evolution of the density, for the above evolution, for $t\in [0,10]$ (d), for $t\in [10,20]$ (e), and $t\in [20,30]$ (f).  
	}
	\label{figure2}
\end{figure}

For $t\in[1.3, 2.5]$, the centered localized waveform exhibits an algebraic in
time growth/decay rate, close to that of the PRW-soliton; notably, both the time-growing, and then
time-decaying, centered profiles appear to manifest
a locking to a PRW-type mode, as it is depicted in the left panel (a) of the top row of Fig~\ref{figure2}, showing the time-evolution of the density of the center $|u(0,t)|^2$, for $t\in[0,3]$. The middle-panel (b) of the top row of Fig~\ref{figure2}, shows a detail of the spatial profile of the maximum event at $t^*=2.415$, close to the right of the two symmetric minima of the exact PRW $u_{\mbox{\tiny PS}}(x,t-2.415;0.84)$. This detail illustrates that the emerged extreme event, preserves the algebraic spatial decay of the PRW soliton.  

The top right panel (c) of Fig. \ref{figure2} depicts a rescaled, extended view of the maximum extreme event, plotted for $x\in [-L,L]$. It reveals a remarkable feature of the dynamics: the extreme event occurs on the top of a finite background of amplitude $|h|^2\sim 1.19$, which is formed {\it at the early stages of the evolution of the vanishing initial condition}: on the one hand,  we observed that the core of the PRW-solitonic structure is proximal to the analytical PRW of the integrable NLS with $P_0=0.84$ (as numerically detected by the fitting argument of \cite{BS} discussed above), and on the other hand, does not tend to the background of unit amplitude as the PRW of the integrable NLS when $\nu=\sigma=1$. This fact, that the amplitude of the background of the PRW-type event differs from $P_0=0.84$ or $P_0=1$, suggests that it may be determined by the driver, as it will be analyzed below. Note that the same effects were observed for increased values of $L$.

The bottom row of Fig.~\ref{figure2} depicts contour plots of the spatiotemporal evolution of the density $|u(x,t)|^2$. The bottom left panel (d) portrays the evolution for $t\in [0,10]$. The PRW-type soliton discussed above, is the first extreme event (FE) corresponding to the spot marked by the arrow. After its formation, the sustaining finite background exhibits modulational instability (MI) dynamics, characterized by the emergence of large amplitude localized modes. A first interesting effect, is that the solution preserves at the early stage of the evolution, the even spatial symmetry $u(x,t)=u(-x,t)$. The even symmetry breaking due to the presence of the driver, occurs 
for $t\gtrsim 14$, as shown in the contour plot of the bottom middle panel (e), which portrays the dynamics for $t\in[10,20]$. A second interesting effect, is that the later stages of MI are manifested by spatial energy localization (initiated prior to the even symmetry breaking), and the formation of extreme amplitude solitary modes. The survived solitary modes are dominating in the dynamics, as shown in the right contour plot (f) (which shows the relevant evolution for $t\in[20,30]$), and their amplitude is increasing. 
\begin{figure}[tbb!]
	\hspace{-0.5cm}\includegraphics[scale=0.21]{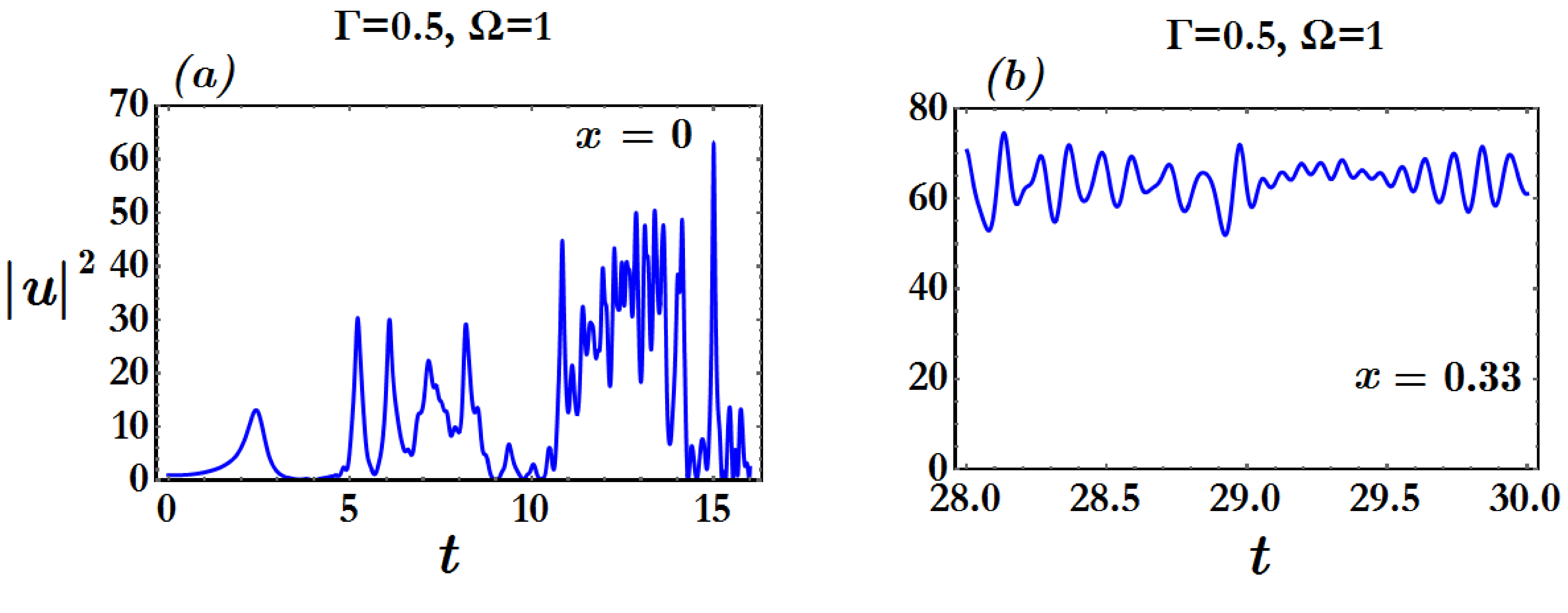}
	\caption{Parameters: $\nu=1$, $\sigma=1$, $\Gamma=0.5$, $\Omega=1$, $L=100$.  Left panel (a): Temporal evolution of the density of the center $|u(0,t)|^2$, for the initial condition (\ref{eq4}). Right panel (b): A detail of the temporal evolution of the density at $x=0.33$ ($|u(0.33,t)|^2$), for  $t\in[28, 30]$.  
	}
	\label{figure2A}
\end{figure}

\begin{figure}[tbb!]
	\hspace{-0.5cm}\includegraphics[scale=0.21]{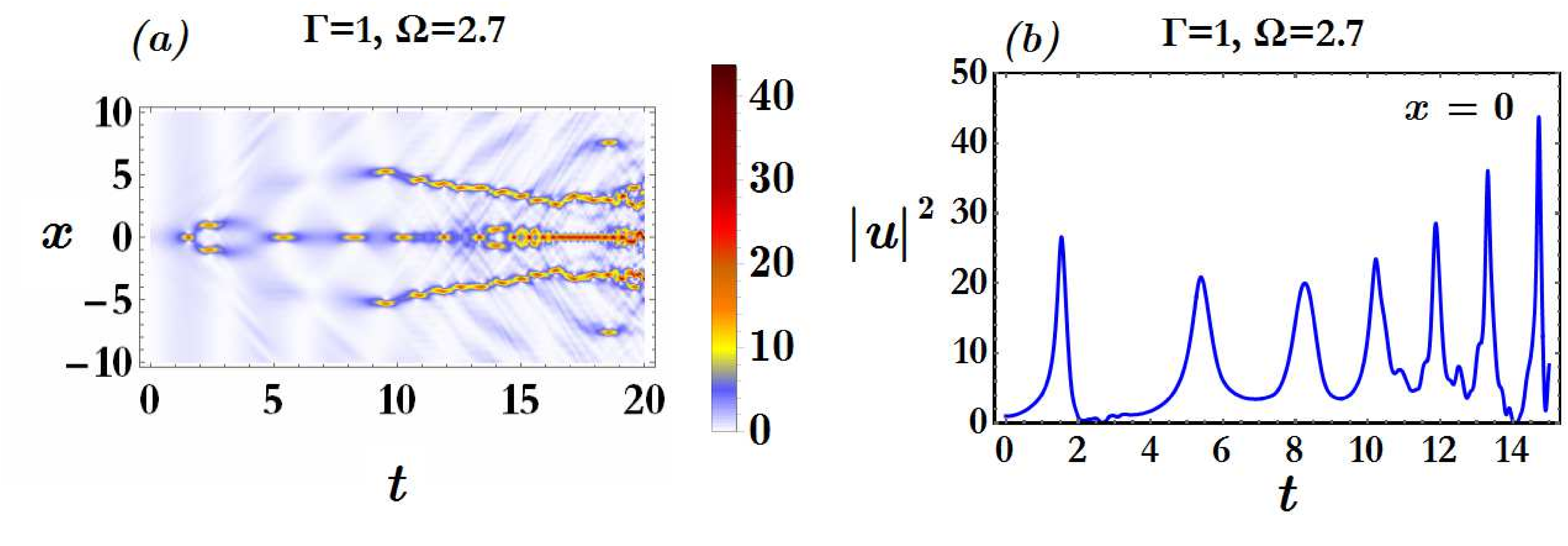}
	\caption{Parameters: $\nu=1$, $\sigma=1$, $\Gamma=1$, $\Omega=2.7$, $L=100$, and initial condition (\ref{eq4}).  Left panel (a):  Contour plot of the spatiotemporal evolution of the density for $t\in [0,20]$. Right panel (b):  Temporal evolution of the density of the center $|u(0,t)|^2$, for $t\in [0,14]$.
	}
	\label{figure2AA}
\end{figure}

Fig. \ref{figure2A} is an attempt to shed light to the structure of the above solitary modes. The left panel (a)  depicts the evolution of the density of the center for $t\in [0,16]$, showing that it undergoes chaotic oscillations. The oscillations in the sub-interval $[10,16]$ correspond to those of the top of the solitary mode which was depicted in the contour plot of the evolution of the density of Fig. \ref{figure2} (e). This is a first evidence that the solitary mode possesses the structure of a (large amplitude) ``chaotic soliton" in the sense of \cite{BorisCS},\cite{nobe1},\cite{nobe2},\cite{NB86}, than a breather. A second evidence is illustrated  in the middle panel (b), showing a a plot of the evolution of the density of $x=0.33$, i.e., $|u(0.33,t)|^2$, for $t\in [28,30]$; it is actually a detail--in this time-subinterval--of the evolution of the solitary mode depicted in the contour plot of Fig. \ref{figure2} (e). The centered mode has slightly slided at $x=0.33$, and the oscillations of the mode are reminiscent of those presented in \cite[Fig.1, pg. 4]{BorisCS}. We may conjecture that the system, for the considered example of parameters is locked to a "chaotic" soliton, and not to a large amplitude breathing mode. Nevertheless, varying the parameters of the driver, the appearance of more breather-like waveforms may not excluded (at least, at the early stages of the evolution), as shown in Fig.
\ref{figure2AA}, depicting the dynamics when $\Gamma=1$ and $\Omega=2.7$. These breather-like modes may evolve to the aforementioned ``chaotic" solitons, at later stages of the dynamics.

\begin{figure}[tbh!]
	\hspace{-0.5cm}\includegraphics[scale=0.21]{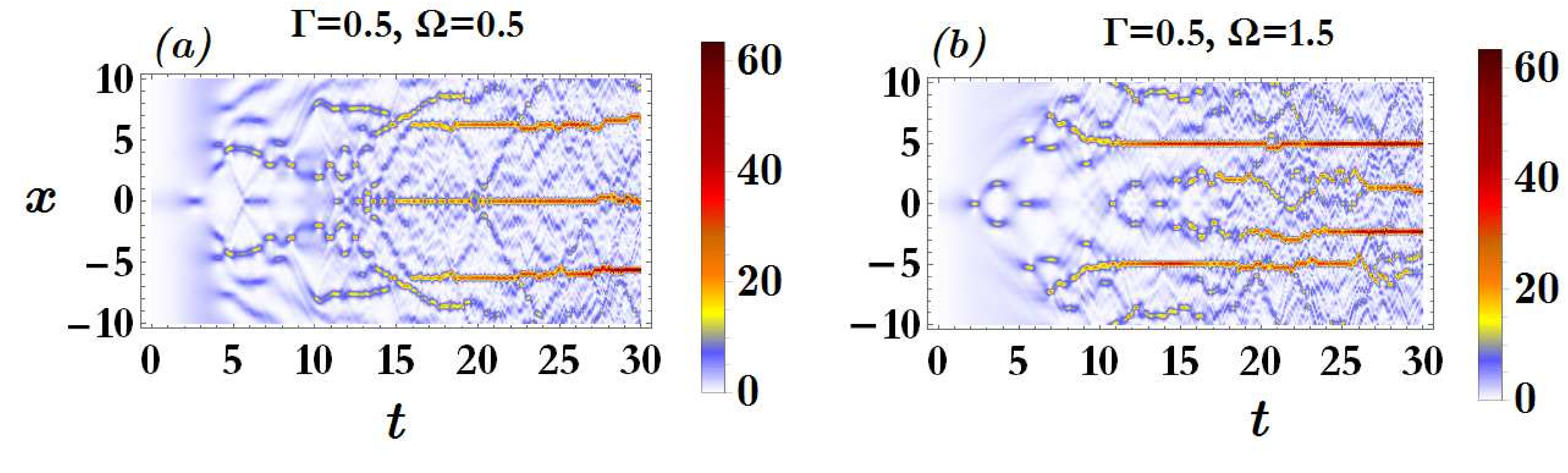}
	\caption{ (Color Online) Parameters: $\nu=1$, $\sigma=1$, $\Gamma=0.5$, $L=100$, and initial condition (\ref{eq4}).  Left panel (a):  Contour plot of the spatiotemporal evolution of the density for $t\in [0,30]$, when $\Omega=0.5$.  Left panel (b):  as in panel (a), but for $\Omega=1.5$.
	}
	\label{figure2B}
\end{figure}

\begin{figure}[tbh!]
	\hspace{-0.5cm}\includegraphics[scale=0.183]{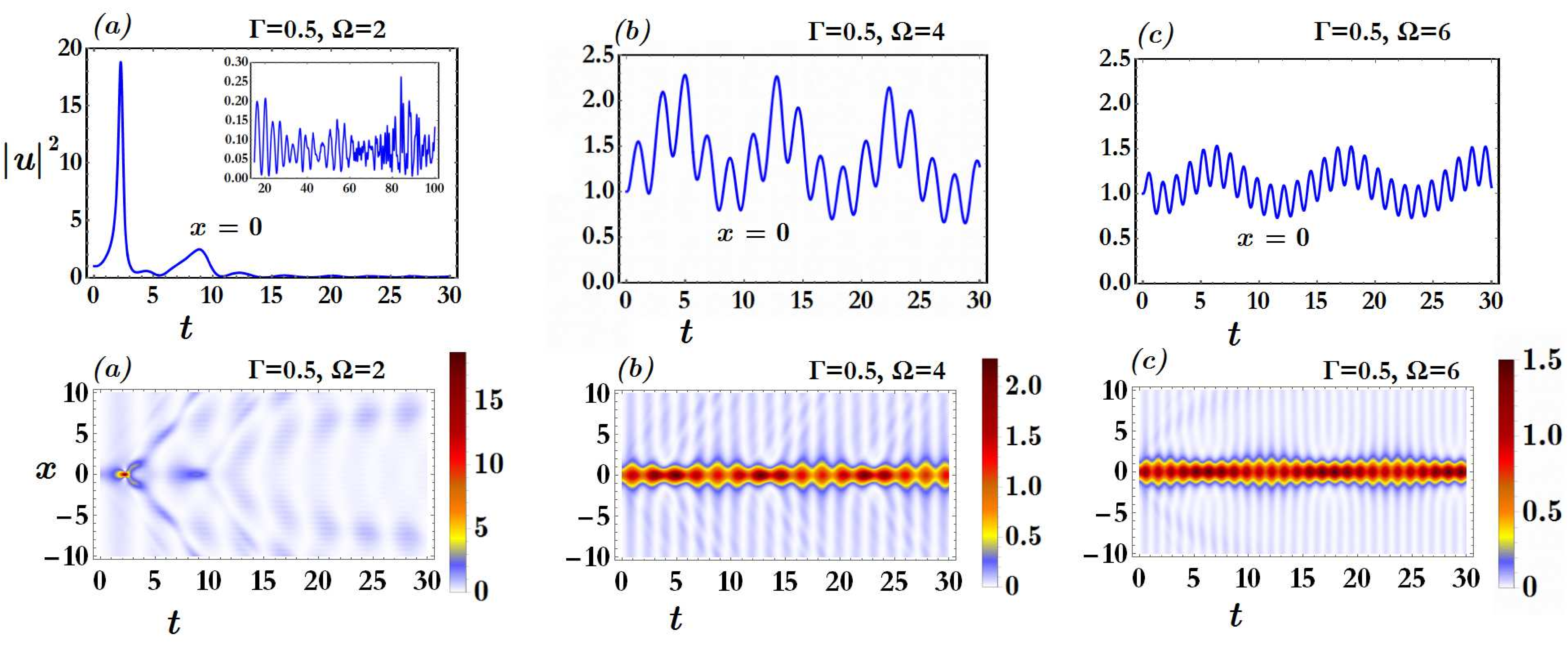}
	\caption{ (Color Online) Parameters: $\nu=1$, $\sigma=1$, $\Gamma=0.5$, $L=100$, and initial condition (\ref{eq4}). Column (a): The upper panel (a) shows the temporal evolution of the density of the center $|u(0,t)|^2$, for $t\in[0,30]$, when $\Omega=2$. The bottom panel (a) shows the contour plot of the spatiotemporal evolution of the density for $x\in [-10,10]$ and $t\in [0,30]$, when $\Omega=2$. Column (b): Same as in column (a), but for $\Omega=4$. Column (c). Same as in column (a), but for $\Omega=6$. 
	}
	\label{figure2C}
\end{figure}

The numerical results that follow, come out from an indicative study on the dependencies of the above dynamics on the parameters of the driver. In Figs. \ref{figure2B} and \ref{figure2C}, we fixed  $\Gamma=0.5$ and $\nu=\sigma=1$ as above, and we varied its frequency $\Omega$. The left panel  (a) of Fig. \ref{figure2B} depicts the contour plot of the spatiotemporal evolution of the density for $\Omega=0.5$ and the right panel (b) for $\Omega=1.5$. Despite some changes in the patterns, the  overall picture of the dynamics observed in the case $\Omega=1$ (large amplitude solitons, following after the emergence of extreme FE), persists for both examples of $\Omega=0.5<1$ and $\Omega=1.5>1$, respectively.  

Drastic changes appear for larger values of the driver's frequency $\Omega$. These changes are illustrated in the columns (a)-(c) of Fig. \ref{figure2C}.  In each column, the upper panel shows the temporal evolution of the density of the center $|u(0,t)|^2$ for $t\in[0,30]$, and the bottom panel shows a contour plot of the spatiotemporal evolution of the density for $x\in [-10,10]$ and $t\in [0,30]$. Column (a) depicts the numerical results for $\Omega=2$, a value which in the present study--for the considered set of parameters-- can be viewed as "critical": The large amplitude peak of the density of the center observed in the top panel (a) corresponds to a PRW-type event--the spot of the bottom panel (a).  Remarkably, afterwards--in contrast with the previous observations-- we see that the large amplitude, ``chaotic solitary'' modes, disappear; the later stages of the dynamics is manifested by small amplitude chaotic oscillations instead, as depicted in the inset of the top panel (a).
Increasing the driver's frequency to $\Omega=4$, we observe in column (b), yet another remarkable effect: the disappearance of the PRW-type events. The dynamics is locked to a spatially localized mode whose top is oscillating in time almost periodically with moderate amplitudes. The frequency of the oscillations of the top of such ``quasiperiodic" solitary modes, seems to be dictated by the frequency of the driver and increasing, as shown in column (c), depicting the relevant evolution for the increased value of $\Omega=6$. When $\Omega$ is further increased, the frequency of the above oscillations is also increasing, suggesting that the dynamics tends to lock to a stationary soliton. This is expected, since in the limit of large $\Omega$, as the period of the oscillations is dictated by the frequency of the driver, should tend to zero.
\begin{figure}[tbh!]
	\hspace{-0.5cm}\includegraphics[scale=0.19]{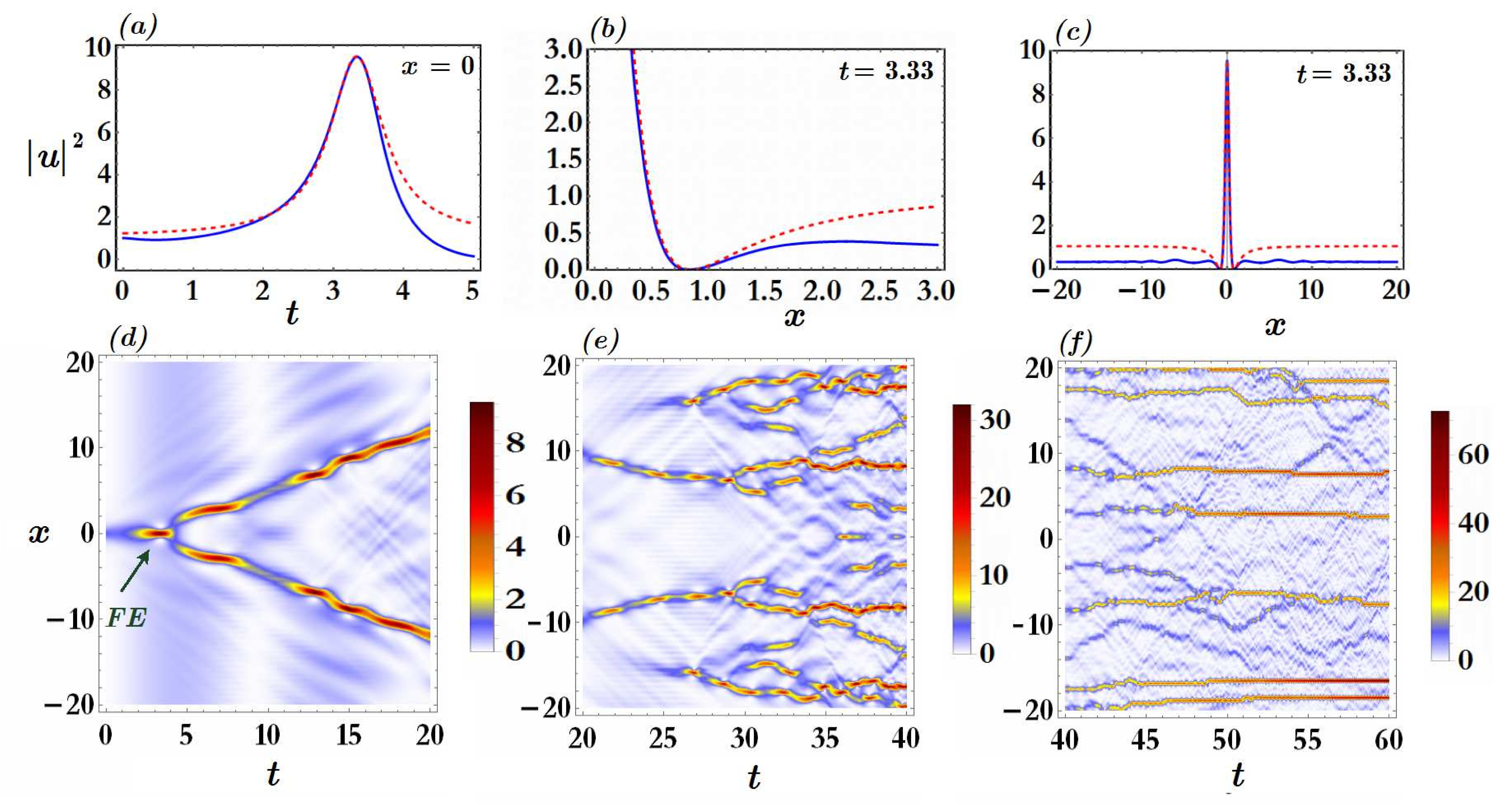}
	\caption{(Color Online) Description  of the dynamics of the initial condition \eqref{eq4}, as in Fig.~\ref{figure2},  but for forcing amplitude $\Gamma=0.25$. In the top panels, the comparison is made against the analytical PRW $u_{\mbox{\tiny PS}}(x,t-3.33;1.06)$,  with $K_0=0.97$ and $\Lambda=0.94$. 
	}
	\label{figure3}
\end{figure}

Next, keeping the driver's frequency fixed to $\Omega=1$, a similar dynamical phenomenology to the one presented in Fig. \ref{figure2} emerged for the  reduced forcing amplitude $\Gamma=0.25$. The dynamics for this example are summarized in Fig. \ref{figure3}. In this case, the FE is found to be close to the PRW-soliton  $u_{\mbox{\tiny PS}}(x,t-3.33;1.06)$ of the integrable limit, with $K_0=0.97$ and $\Lambda=0.94$. The presentation is the same as in Fig. \ref{figure2}. The top left panel (a) shows the time evolution of the density of the center, where its time-growth and time decay is still close to the PRW for $t\in[2,3.5]$. The top middle-panel (b), illustrates that the numerical solution, when the FE attains it maximum density at $t^*=3.33$, yet captures the profile of the PRW around its symmetric minima, even closer than the case of $\Gamma=0.5$. The whole profile of the FE at the time of its maximum density, is depicted in the top right-panel (c). As a result of the decreased forcing amplitude, the amplitude of the finite background supporting the PRW-event is also decreased, i.e., $|h|^2\sim 0.34$. Accordingly, the emerging localized modes posses reduced amplitude. Additionally, we observe a delay in the emergence of the extreme amplitude solitary modes, as shown in the panels (d)-(e)-(f) of the bottom row, portraying contour plots of the spatiotemporal evolution of the density, for $t\in [0,60]$.
\begin{figure}[tbh!]
	\hspace{-0.5cm}\includegraphics[scale=0.19]{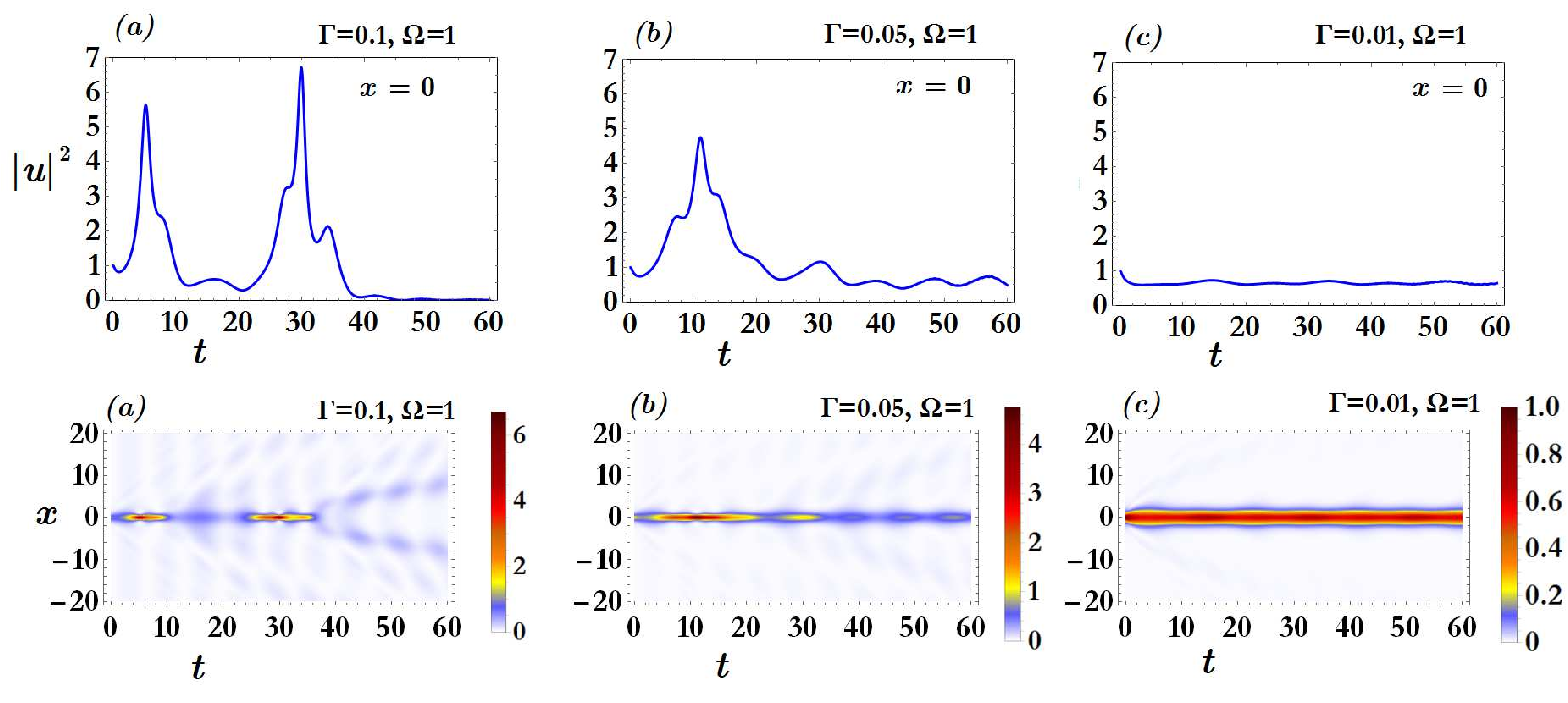}
	\caption{(Color Online) Parameters: $\nu=1$, $\sigma=1$, $\Omega=1$, $L=100$, and initial condition (\ref{eq4}). Column (a): The upper panel (a) shows the temporal evolution of the density of the center $|u(0,t)|^2$, for $t\in[0,30]$, when $\Gamma=0.1$. The bottom panel (a) shows the contour plot of the spatiotemporal evolution of the density for $x\in [-10,10]$ and $t\in [0,30]$, when $\Gamma=0.1$. Column (b): Same as in column (a), but for $\Gamma=0.05$. Column (c): Same as in column (a), but for $\Gamma=0.01$.  }
	\label{figure3A}
\end{figure}


Proceeding to a progressive decrease of $\Gamma$, we observe a suppression of the extreme wave dynamics (similar to the case of increasing $\Omega$). These suppression effects are illustrated in Fig. \ref{figure3A}, where the presentation follows that of Fig. \ref{figure2C}: for $\Gamma=0.1$, we observe in column (a), that first, the large amplitude solitary structures disappear while the emergence of  rogue waves  still persists. This feature is shown by the large amplitude peaks of the density of the center shown in the top panel (a), which correspond to the localized spots of the contour plot portrayed in the bottom panel (a). Further suppression occurs for $\Gamma=0.05$ as shown in column (b) (manifested by the decrease of amplitude of the FE), while for $\Gamma=0.01$, the dynamics seems again, to tend to lock to a stationary soliton.

A complete study of the bifurcations in the full parameter $(\Gamma,\Omega)$-parametric space, apart of being essential, it might be a formidable task (as it may involve the non-trivial analysis of resonances given in \cite{Rev1},\cite{RK1},\cite{RK2}), and is beyond the scope of the present work. However, we may already conjecture on the dependencies of the exhibited dynamics on the amplitude $\Gamma$ and the frequency $\Omega$ of the driver. For instance, for fixed $\nu$, $\sigma$, $L$,  we may identify thresholds $\Gamma_{\mathrm{thresh}}$ and $\Omega_{\mathrm{thresh}}$ such that for suitably fixed $\Omega$ (or $\Gamma$), if $\Gamma>\Gamma_{\mathrm{thresh}}$ (or $\Omega<\Omega_{\mathrm{thresh}}$),  extreme wave dynamics emerge.  The above numerical studies provided the following examples for the thresholds:  we found that for $\nu=\sigma=1$ and $L=100$, when $\Omega=1$, $\Gamma_{\mathrm{thresh}}<0.05$,  and when $\Gamma=0.5$, then $\Omega_{\mathrm{thresh}}>2$. Furthermore, in the suppression regimes, in the limit of small $\Gamma$ (for fixed $\Omega$) or in the limit of large $\Omega$ (for fixed $\Gamma$), the dynamics tends to lock to a stationary soliton of the integrable NLS. This is expected, as in the limit of small $\Gamma$, the system approximates the integrable limit.
	
The above observations will be further underlined by the  comments on the behavior of the integrable limit $\Gamma=0$, for the same type of vanishing conditions. 
\paragraph{Effects from continuous wave solutions.}
The existence and stability properties of continuous wave (cw) solutions of the forced NLS Eq. \eqref{eq1}, should have an important role on the birth of the transient PRW-type dynamics. In what follows, we fix for simplicity $\nu=\sigma=1$ in \eqref{eq1}, corresponding to the presented results of the numerical simulations discussed above. We consider spatially homogeneous solutions, of the form 
\begin{equation}\label{eq6}
u(x,t) = he^{\mathrm{i}\Omega t}, \quad h\in\mathbb{C}.
\end{equation}
There exist for Eq. \eqref{eq1}, under the dispersion relation
\begin{equation}\label{eq7}
|h|^2h = \Gamma +h\Omega.
\end{equation}
For $\Gamma>0$,  the case we are restricted herein, Eq.~\eqref{eq7} has only real solutions for $h$:  Indeed, let $h=A+\mathrm{i}B$. Then,  its substitution to \eqref{eq7}, is leading to the following equations for $A$ and $B$:
\begin{eqnarray*}
A(A^2+B^2)-\Omega A&=&\Gamma,\\
B(A^2+B^2)-\Omega B&=&0.
\end{eqnarray*}
From the second equation, we have that either $B=0$, or $|h|^2=A^2+B^2=\Omega$. Inserting the latter, to the first equation for $A$, we get that $\Gamma=0$. Thus, for $\Gamma>0$ (and  generically, for $\Gamma\in\mathbb{R}$), solutions \eqref{eq6} exist only with $h\in\mathbb{R}$.
Now, consider a  perturbation to the solution \eqref{eq6}, of the form
\begin{equation}
\label{eq8a}
u(x,t)=[h + \epsilon u_1(x,t)]e^{\mathrm{i}\Omega t},
\end{equation}
for small $\epsilon>0$, which we insert into  Eq.~\eqref{eq1}. By using the dispersion relation \eqref{eq7}, and linearizing the system, i.e. neglecting terms of order $\epsilon^2$ and higher, we derive the equation this small  influence satisfies:
\begin{eqnarray}
\label{eq8b}
\mathrm{i}\partial_tu_1+\frac{1}{2}\partial^2_{x}u_1+2h^2u_1+h^2\overline{u_1}-\Omega u_1=0.
\end{eqnarray}
To examine MI, we may assume that the perturbation $u_1$ is harmonic, i.e.
\begin{equation}
u_1(x,t)=c_1 e^{i (k x-\omega  t)} + c_2 e^{-i (k x-\omega  t)},
\label{eq9}
\end{equation}
where $k$ and $\omega$ denote the wavenumber and the frequency of the perturbation, respectively. Next, substitution of the expression \eqref{eq9} for $u_1$ in the linearized equation \eqref{eq8b}, yields the following algebraic system for $c_1$ and $c_2$:
\begin{eqnarray*}
&&\left(-\frac{k^2}{2}+\omega-\Omega +2h^2\right)c_1+ h^2c_2= 0,\\
&& h^2c_1 +\left(-\frac{k^2}{2}-\omega-\Omega+2h^2\right)c_2= 0.
\end{eqnarray*}
Seeking for nontrivial solutions $c_1$ and $c_2$ of the above system, we require the
relevant determinant to be zero; this way, we obtain the following dispersion relation:
\begin{eqnarray}
\label{eq10}
\omega^2=\frac{k^4}{4}+k^2(\Omega-2h^2)+(3h^4-4\Omega h^2+\Omega^2).
\end{eqnarray}
The pertubation \eqref{eq8a}, suggests that solutions \eqref{eq6} are modulationally unstable if $\omega$ is complex, i.e., when the right-hand side of \eqref{eq10} is negative. Solving the equation  
\begin{eqnarray*}
k^4+k^2\left(4\Omega-8h^2\right)+12h^4-16\Omega h^2+4\Omega^2=0,
\end{eqnarray*}
in terms of $k$, we find the following solutions:
\begin{eqnarray}
\label{eq11}
k_{1,2}=\pm \sqrt{2}\sqrt{h^2-\Omega},\;\;k_{3,4}=\pm \sqrt{2}\sqrt{3h^2-\Omega}.
\end{eqnarray}
The roots \eqref{eq11} define the instability bands of the cw solutions \eqref{eq6}. There is not a loss of generality to be restricted for $k>0$.  Then, the instability bands are defined as follows:
\begin{enumerate}
\item $I_{a}=[0,k_3]=[0, \sqrt{2}\sqrt{3h^2-\Omega}]$, if $\frac{\Omega}{3}<h^2<\Omega$.
\item $I_{b}=[k_1,k_3]=[\sqrt{2}\sqrt{h^2-\Omega}, \sqrt{2}\sqrt{3h^2-\Omega}]$, if $h^2>\Omega$.
\end{enumerate}

MI has been proved an essential mechanism for the emergence of rogue waves 
\cite{EPeli},\cite{Kharif1},\cite{Kharif2}. In the light of the  MI analysis recalled above, let us reconsider the dynamics presented in the example of Fig.~\ref{figure1} and Fig.~\ref{figure2}. When $\Omega=1$ and $\Gamma=0.5$, equation  \eqref{eq7} has one real root
$h_1 = 1.19$. Thus,  the  corresponding cw solution \eqref{eq6}, with $h=h_1$ and exhibits its  modulational instability for wave numbers $k\in I_{b}$, since $h_1^2>\Omega$. 

In this regard, we may conjecture that the background sustaining the PRW-type waveform shown in the snapshots of Fig.~\ref{figure1} (and panel (c) of Fig. \ref{figure2}), is self-induced as the system tends transiently to lock  its cw solution in the presence of the forcing, e.g, the solution of amplitude $|h_1|^2$ in the considered example. However, this solution is modulationally unstable.  Then, as a  transient ``metastability effect'', the emergence of the PRW-type waveform is a result of the synergy of the preservation of the spatial localization of the initial condition (due to continuous dependence on the initial data \cite{All2}), and of the MI of the sustaining cw-solution:  at a fixed time (recall the snapshot at time $t=2$ in Fig.~\ref{figure1}), a pulse on a finite background has formed, although the evolution initiated by vanishing initial conditions; the system  has self-induced the  universal effects of the MI mechanism analyzed in \cite{BM},\cite{Yang1},\cite{Yang2}, which may lead to the birth of a PRW-type mode. The same mechanism may explain the dynamics portrayed in Fig.~\ref{figure3}. 

We should remark that the above arguments are further  supported by the detailed analysis of \cite{Rev1}, on the adiabatic excitation and control of $N$-band solutions ($N$-phase waves) for the forced NLS. Particularly relevant is the analysis on the excitation of the spatially homogeneous ($0$-band) solutions \eqref{eq6} from {\em zero initial conditions}, which are continuously synchronized with the driver (despite the variation of the driver's frequency); in our case the vanishing initial conditions define perturbations of the zero background (being modulationally stable in the case of the integrable limit $\Gamma=0$).  
\begin{figure}[tbh!]
	\hspace{-0.5cm}\includegraphics[scale=0.19]{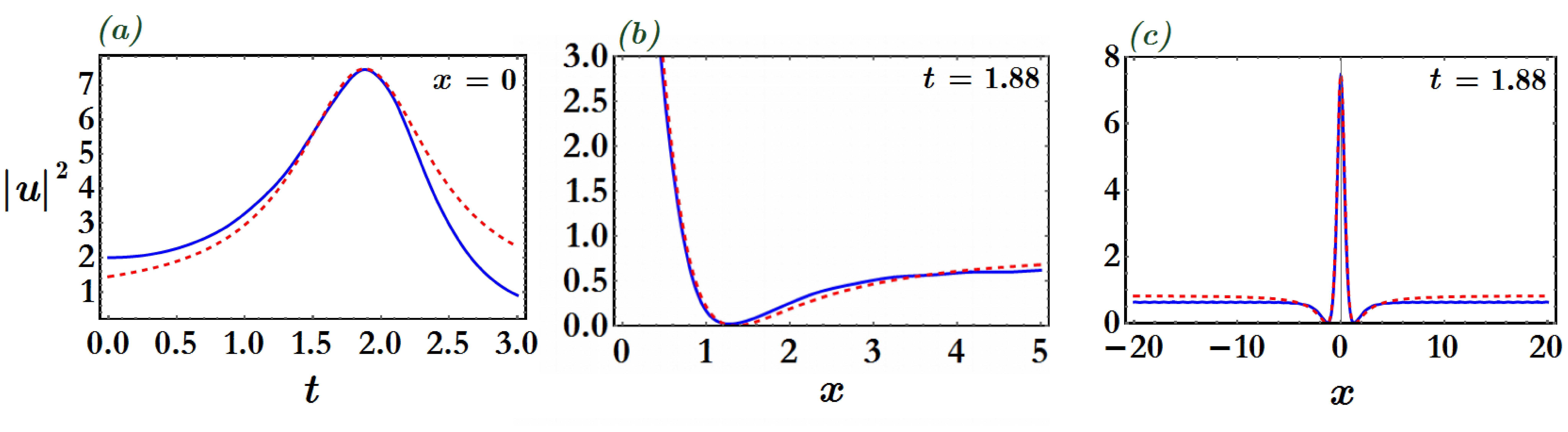}
	\caption{(Color Online)  Description  of the dynamics of the initial condition \eqref{eq5} with $\alpha=\sqrt{2}$ and $\beta=1$, as in Fig.~\ref{figure2},  but for the damped and forced NLS Eq.~\eqref{eq1d}. Parameters: $\nu=2$, $\sigma=1$, $\gamma=0.02$, $\Gamma=0.5$, $\Omega=1$, $L=100$. The comparison is made against the analytical PRW $u_{\mbox{\tiny PS}}(x,t-1.88;0.83)$, with $K_0=1.55$ and $\Lambda=1.20$. 
	}
	\label{figure4}
\end{figure}
\paragraph{Comment on the dynamics of the damped counterpart.}
The  effects of linear loss, solely influencing the evolution of Peregrine solitons in the $1$D-focusing NLS have been analyzed via nonlinear spectral analysis, in \cite{Rev1_c}; the unforced, damped NLS equation is physically significant in hydrodynamics and nonlinear optics \cite{Kharif1},\cite{Kharif2},\cite{Rev1_d}. Numerical and experimental studies \cite{Rev1_e}, confirmed the observation of higher-order MI dynamics in water waves.  

Here, we illustrate that dynamical behavior of \eqref{eq1} discussed in the previous paragraphs seems to be robust for small damping strengths in the presence of the periodic forcing. For instance, this robustness was identified for  the linearly damped counterpart of \eqref{eq1}: 
\begin{eqnarray}
\label{eq1d}
\mathrm{i}{{u}_{t}}+\frac{\nu}{2}{{u}_{xx}}+\sigma|u|^2u+\mathrm{i}\gamma u =\Gamma\exp(\mathrm{i}\Omega t),\;\;\gamma>0.
\end{eqnarray}
Yet, this model is of particular interest in various physical contexts, as in plasma physics \cite{BoYu}, \cite{NB86} (governing the dynamics of a collisional plasma driven by an external rf field).  The dynamics of Eq.~\eqref{eq1d}, are captured by a finite dimensional global attractor. For its existence and analyticity properties, we refer to \cite{Ghid88},\cite{Goubet1}.  

Fig.~\ref{figure4} depicts the results of the numerical study for $\nu=2$, $\sigma=1$, damping strength $\gamma=0.02$ and $\Gamma=0.5$. This time, we have used the $\mathrm{sech}$-profiled initial condition \eqref{eq5}, for $\alpha=\sqrt{2}$ and $\beta=1$ corresponding to an exact stationary pulse of the integrable NLS (see below). As before, the left panel (a) shows the time evolution of the density of the center $|u(0,t)|^2$, the middle panel (b) shows the detail of the emerged PRW-type event around the right of its minima, and the right panel (c) the profile of the PRW-type event when its maximum density is attained at time $t^*=1.88$.  The comparison is against the analytical PRW solution \eqref{sprw} $u_{\mbox{\tiny PS}}(x,t-1.88;0.83)$ with $K_0=1.55$ and $\Lambda=1.20$. Again, the excitation of the PRW-type waveforms can be explained as above, in terms of the modulation instability of the cw-solutions \eqref{eq6} of Eq. \eqref{eq1d} (see \cite{NB86}). Additionally, it would be interesting to examine  a potential stabilization of the solitary modes observed in Fig. 2 (e) and (f), in the presence of small damping, when possibly driven by two frequencies, as proposed in \cite{Malf}.
\begin{figure}[tbh!]
	\hspace{-0.5cm}\includegraphics[scale=0.19]{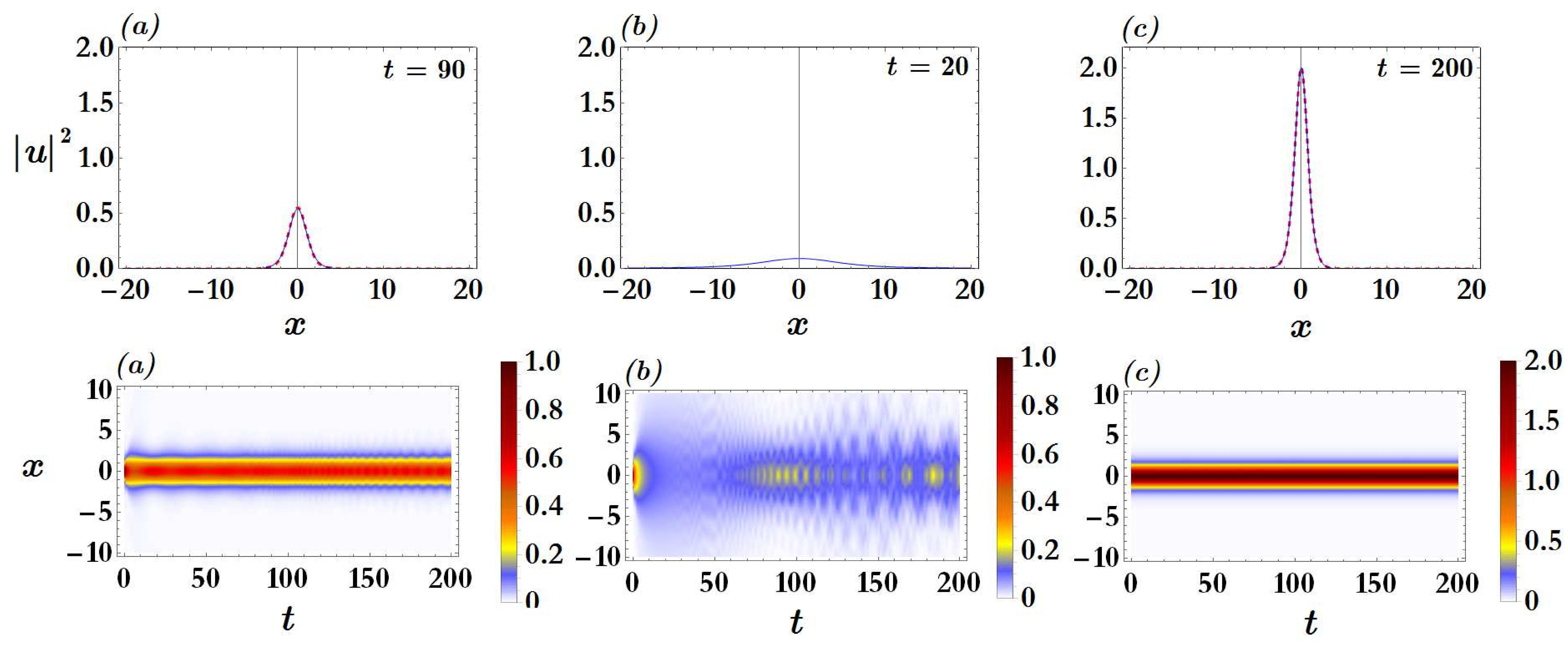}
	\caption{(Color Online) Left column (a):  Parameters in Eq.~\eqref{eq1}, $\nu=\sigma=1$, $\Gamma=0$. The top panel (a) shows a snapshot of the numerical density of the solution with initial condition \eqref{eq4}, at $t=90$ [continuous (blue) curve], against the density of an exact stationary soliton $|\Phi|^2\approx 0.54\,\mathrm{sech}^2(0.73x)$ [dashed (red) curve]. The bottom panel (a) shows a contour plot of the spatiotemporal evolution of the density, for parameters and initial condition as in the top panel (a).  Middle column (b): Parameters in Eq.~\eqref{eq1}, $\nu=2$, $\sigma=1$, $\Gamma=0$. The top panel (b) shows a snapshot of the density of the numerical solution for the initial condition \eqref{eq4}, at $t=20$.  The bottom panel (b) shows a contour plot of the spatiotemporal evolution of the density, for parameters and initial condition as in the top panel (b).  Right column (c): Parameters in Eq.~\eqref{eq1}, $\nu=2$, $\sigma=1$, $\Gamma=0$. The top panel (c) shows a snapshot at $t=200$, of the density of the  numerical solution [continuous (blue) curve], with initial condition \eqref{eq5} ($\alpha=\sqrt{2}$, $\beta=1$), against the density of the corresponding analytical stationary soliton [dashed (red) curve]. The bottom panel (b) shows  a contour plot of the spatiotemporal evolution of the density for parameters and initial condition as in the top panel (c). In all studies $L=100$. 
	}
	\label{figure5}
\end{figure}
\paragraph{Comment on the dynamics of the integrable NLS limit.}
It is important to note that the dynamics exhibited by the integrable NLS assuming the initial condition  \eqref{eq4} or \eqref{eq5}, totally differs from those of the forced (and damped) counterpart discussed in the previous paragraphs, though well understood. Fig. \ref{figure5} summarizes the results of the numerical study of Eq.~\eqref{eq1}, for $\Gamma=0$. 

The left column (a) depicts numerical results for the evolution of the initial condition \eqref{eq4}, when $\nu=\sigma=1$. The upper panel (a) shows a snapshot of the density of the numerical solution at time $t=90$ [(blue) continuous curve], against the density of a numerically detected stationary pulse $|\Phi|^2\approx 0.54\,\mathrm{sech}^2(0.73x)$ (dashed red curve). Let us recall the standard formula of the general bright-soliton solution of the  cubic integrable NLS:
\begin{eqnarray}
\label{sw}
u(x,t)=\sqrt{2\left|\frac{\theta-\kappa^2\nu/2}{\sigma}\right|}\mathrm{sech}\left[\sqrt{\left|\frac{\theta-\kappa^2\nu/2}{\nu/2}\right|}(x-2\theta\kappa t)\right]\mathrm{exp}[\mathrm{i}(\kappa x-\theta t)].
\end{eqnarray}
When $\sigma=\nu=1$ and $\kappa=0$, the corresponding stationary solution has density $|u|^2=2|\theta|\mathrm{sech}^2\left(\sqrt{2|\theta|}x\right)$. Setting $2|\theta|=0.54$, we deduce that indeed, the numerically detected $|\Phi|^2$ corresponds to the density of an exact stationary pulse.
The approaching  of $|\Phi|^2$ is illustrated in the contour plot of the spatiotemporal evolution of the density, shown in the bottom left panel (a). We observe the slight oscillations of the numerical solution around $\Phi$. This oscillatory motion is completely determined by the well known stability results of \cite[Theorem 8.3.1]{Caz03} for standing wave solutions, if applied in the case of the cubic NLS: the requested distance between the initial condition \eqref{eq4} and $\Phi$ is quantitatively significant in the Sobolev norm $H^1(\mathbb{R})$, i.e., $\|u_0-\Phi\|^2=\int_{-\infty}^{\infty}\left|u_0-\Phi\right|^2dx+\int_{-\infty}^{\infty}\left|\partial_xu_0-\partial_x\Phi\right|^2dx\,\approx 0.97$. Due to this distance, the excited numerical solution, although it seems to be almost locked to $\Phi$, exhibits the observed oscillatory behavior.

The middle column (b) depicts the dynamics of the initial condition \eqref{eq4}, for $\nu=2$, $\sigma=1$. The initial condition disperses, as shown in the snapshot of the density for $t=20$ of the upper panel (b), and the contour plot of its spatiotemporal evolution, shown in the bottom panel (b). 

Finally, column (c) depicts the dynamics of the initial condition \eqref{eq5}, for $\alpha=\sqrt{2}$ and $\beta=1$, when $\nu=2$ and $\sigma=1$. In this case, the initial condtion $u_0(x)=\sqrt{2}\mathrm{sech}(x)$ corresponds to the exact stationary pulse \eqref{sw} at $t=0$, for $\kappa=0$, $\theta=1$, and the above values of $\nu$ and $\sigma$. The snapshot of the numerical density [(blue) continuous curve], at $t=200$, shown in the upper right panel (c), fits exactly to the density of the exact stationary solution [(red) dashed curve]. The contour plot of the spatiotemporal evolution of the density is an illustration of the well known stability of the exact soliton pulse.

It should be also highlighted, that for the generalized focusing Hamiltonian NLS 
\begin{eqnarray*}
\label{eq1g}
\mathrm{i}{{u}_{t}}+\frac{\nu}{2}{{u}_{xx}}+\sigma|u|^{2\delta}u =0,\;\;\delta\geq 1, 
\end{eqnarray*}
($\delta=1$ corresponds to the integrable NLS), rogue waves can be still excited by spatially decaying initial conditions as it was found in \cite{BS}. However, the observed extreme waves therein (excited by generic Gaussian wave packets as their width is varied) are decaying to zero. This is a vast difference of the results of \cite{BS}, with those presented in the present paper.

Summarizing, comparing the dynamics of the integrable NLS ($\Gamma=0$), with those of the forced ($\Gamma > 0$)  Eq.~\ref{eq1} [and damped ($\gamma >0$)-Eq.~\eqref{eq1d}], it is clear that the birth of extreme events for the latter, initiated by vanishing initial conditions, is far from any integrable limit approximation, \cite{onorato,brunetti}, and further justifies the potential existence of thresholds for the driver's amplitude and frequency, with the properties described at the end of Section II.c. 


\section{Conclusions}
In this work, direct numerical simulations revealed the excitation of Peregrine-type solitonic waveforms, from vanishing initial conditions (possessing an algebraic or exponential spatial decaying rate), for the periodically driven nonlinear Schr\"odinger (NLS) equation. The PRW-type waveforms emerge as first events of the evolution, on the top of a self-induced finite background. This  dynamical behavior can be understood in terms of the existence and  modulation instability of the continuous wave solutions of the model, and the preservation of the spatial localization of the initial condition at the early stages of the evolution. Revisiting the dynamics of the corresponding conservative NLS for the same type of initial conditions, it was shown that the above dynamics should be considered as far from  approximations from the integrable limit.  We also commented that this behavior may persists in the linearly damped and forced counterpart, at least under the presence of small damping strengths.  Importantly,  it appears that the emergence of the Peregrine soliton excited by decaying initial conditions as a universal, coherent structure in the dynamics of the $1$D--integrable NLS \cite{Rev1_b}--as studied therein, in its semiclassical limit scenario \cite{BM1},\cite{BM2}--can be robust in the presence of forcing and damping. Notably, for the persistence of semiclassical type dynamics in the presence of a spatiotemporally localized driver (pending on the spatial/temporal scales of the latter and the magnitude of the damping strength), we refer to our recent work \cite{LetArx1}.

Future directions, include further investigations on forced and damped NLS models which may be considered in $1\mathrm{D}$ and higher dimensional set-ups, the consideration of various types of forcing (as spatiotemporally localized \cite{LetArx1}), the presence of higher order effects, as well as, discrete \cite{KevreDNLS}, damped and forced NLS counterparts. Relevant investigations are in progress and will be considered in future publications.
\label{conclusions}
\section*{Acknowledgments}
The authors gratefully acknowledge the support of the grant MIS 5004244 under the action $\mathrm{E\Delta BM34}$, funded by European Social Fund (ESF) and Hellenic General Secretariat of Research and Technology (GSRT).\\ 


\end{document}